\catcode`\@=11
\def\gsim{\ifmmode{\mathrel{\mathpalette\@versim>}}
    \else{$\mathrel{\mathpalette\@versim>}$}\fi}
\def\lsim{\ifmmode{\mathrel{\mathpalette\@versim<}}
    \else{$\mathrel{\mathpalette\@versim<}$}\fi}
\def\@versim#1#2{\lower 2.9truept \vbox{\baselineskip 0pt \lineskip 
    0.5truept \ialign{$\m@th#1\hfil##\hfil$\crcr#2\crcr\sim\crcr}}}
\catcode`\@=12

\def\etal{et al.$\,$}
\def\rhos{\rho _*}

\def\rc{r_{\rm c}}
\def\rch{r_{\rm h}}
\def\rcs{r_{\rm *}}
\def\rref{R_{\rm e}}
\def\rst{r_{\rm st}}
\def\ms{M_*}
\def\mh{M_{\rm h}}
\def\msun{M_{\odot}}
\def\min{M_{\rm in}}
\def\mou{M_{\rm out}}
\def\rap{{\cal R}}
\def\phit{\phi_{\rm T}}
\def\phis{\phi_*}
\def\phih{\phi_{\rm h}}
\def\iss{I_{**}}
\def\ish{I_{\rm *h}}
\def\kss{K_{**}}
\def\ksh{K_{\rm *h}}
\def\uss{U_{**}}
\def\ush{U_{\rm *h}}
\def\lx{L_{\rm X}}
\def\lb{L_{\rm B}}
\def\lsun{L_{\odot}}
\def\lsn{L_{\rm SN}}
\def\lgm{L_{\rm g}}
\def\lsi{L_{\sigma}}
\def\ss{\sigma_*}
\def\sgc{\sigma_{\rm c}}
\def\SX{\Sigma_{\rm X}}
\def\vt{\vartheta _{\rm SN}}
\def\Rsn{R_{\rm SN}}
\def\Esn{E_{\rm SN}}
\def\chil{\chi_{\ell}}
\def\es{{\rm erg}\;{\rm s}^{-1}}
\def\ks{{\rm km}\;{\rm s}^{-1}}
\def\apj{ApJ}
\def\apjl{ApJL}
\def\apjs{ApJS}
\def\aj{AJ}
\def\aap{A\&A}
\def\mnras{MNRAS}
\def\araa{ARAA}
\def\pasj{PASJ}
\documentstyle{l-aa}

\begin{document}

   \thesaurus{(11.05.1; 11.09.4; 11.11.1; 13.25.2)}

   \title{Decoupled hot gas flows in elliptical galaxies}

   \subtitle{}

   \author{S. Pellegrini\inst{1} and L. Ciotti\inst{2}} 

   \offprints{S. Pellegrini; e-mail pellegrini@astbo3.bo.astro.it}

   \institute{\inst{1}Dipartimento di Astronomia, Universit\`a di Bologna,
              via Zamboni 33, 40126 Bologna, Italy\\
              \inst{2}Osservatorio Astronomico di Bologna,
              via Zamboni 33, 40126 Bologna, Italy}

   \date{Received ... ; accepted ...}

   \maketitle

   \begin{abstract}

We present the results of a new set of hydrodynamical simulations of
hot gas flows in model elliptical galaxies with the following
characteristics: the spatial luminosity distribution approaches a
power law form at small radii, in accordance with the results of
recent ground based observations, and with the Hubble Space Telescope;
the dark matter has a peaked profile too, as indicated by high
resolution numerical simulations of dissipationless collapse; the
dark to luminous mass ratio spans a large range of values, including
low values found by optical studies confined to within two effective
radii; finally, the type Ia supernova rate is that given by the latest
estimates of optical surveys, or zero, as suggested by the iron
abundances recently measured in the hot gas.

We find that the resulting gas flows are strongly decoupled: an inflow
develops in the central region of the galaxies, while the external
parts are still degassing, i.e., the flows are mostly partial
winds. This behavior can be explained in terms of the local energy
balance of the hot gas. A large spread in the X-ray luminosity $\lx$
at fixed optical luminosity $\lb$ can be produced as in
previous simulations that used King models plus massive
quasi-isothermal dark halos, and higher supernova rates; the key
factor causing large $\lx$ variations is now the size of the central
inflow region. The highest $\lx$ observed correspond to global
inflows. Finally, non negligible amounts of cold gas can be produced by
the partial winds; this could be an explanation for the possible
discovery of cold matter at the center of elliptical galaxies, an
alternative to the presence of a steady state cooling flow.


   \end{abstract}


\section{Introduction}

X-ray observations, beginning with the $Einstein$ Observatory, have
demonstrated that normal early-type galaxies are X-ray emitters, with
0.2--4 keV luminosities ranging from $\sim 10^{40}$ to $\sim
10^{43}\;\es$ (\cite{fab89}; Fabbiano, Kim, \& Trinchieri 1992). The
X-ray luminosity $\lx$ is found to correlate with the blue luminosity
$\lb$ ($\lx\propto \lb^{2.0\pm 0.2}$), but there is a large scatter of
roughly two orders of magnitude in $\lx$ at any fixed $\lb >3\times
10^{10}\lsun$ (Fig.~1). The observed X-ray spectra of galaxies with
high $\lx/\lb$ ratios are consistent with thermal emission from hot,
optically thin gas, while those of low $\lx/\lb$ objects can be mostly
accounted for by emission from stellar sources (Kim, Fabbiano, \&
Trinchieri 1992).

The scatter in the $\lx -\lb$ diagram has been recognized as the most
striking feature of the X-ray properties of early-type galaxies. Using
new apparent magnitudes and fundamental plane distances, it was shown
that this scatter is reduced by 20\%, but not eliminated (Donnelly,
Faber \& O'Connell 1990; see Fig.~1b). This result is based on the old
estimate of the X-ray fluxes (that of Canizares, Fabbiano, \&
Trinchieri 1987), and on just half of the final sample of X-ray
galaxies produced by Fabbiano et al. 1992 (see Eskridge, Fabbiano \&
Kim 1995 for a detailed comparison of the statistical results by
Donnelly et al. with those obtained using the whole sample). One can
argue that the large dispersion in $\lx $ is definitively not the
result of distance errors on the basis of the fact that a scatter of
the same size as in Fig.~1a is present even in the
distance-independent diagram of $\lx/\lb$ versus the central stellar
velocity dispersion (e.g., Eskridge et al. 1995).

Many theoretical models were developed to explain the findings above,
including numerical simulations of the behavior of gas flows fed by
stellar mass loss and heated by type Ia supernovae (SNIa). Steady
state cooling flow models were investigated first (Nulsen et al.
1984, \cite{sw87},1988),
and it was found that these can only reproduce X-ray bright
galaxies. Evolutionary models with a SNIa rate approximately constant
with time have been carried out by Mathews \& Loewenstein (1986),
Loewenstein \& Mathews (1987), and David, Forman, \& Jones
(1990, 1991). After a very brief initial wind phase, driven by the
explosion of type II supernovae, the resulting flow evolution goes
from a global inflow to a wind, which is experienced only by the
smallest galaxies by the present time. David \etal (1991) conclude
that all galaxies above $\lb\simeq 3\times 10^{10}\lsun$ host a
cooling flow. So, as in the steady state cooling flow scenario, the
scatter in $\lx $ at fixed $\lb$ has to be explained by a combination
of environmental differences (\cite{ws91}) and by large variations
from galaxy to galaxy in the stellar mass loss rate per unit $\lb$,
the efficiency of thermalization of the stellar mass loss, and that of
thermal instabilities in the hot gas. An alternative way of explaining
the scatter in the $\lx -\lb$ diagram is given by the evolutionary
scenario of D'Ercole \etal (1989), and Ciotti \etal (1991,
CDPR). Assuming that the SNIa explosion rate is declining with time
slightly faster than the rate with which mass is lost by the stars, in
the beginning the energy released by SNIa's can drive the gas out of
the galaxies through a supersonic wind. As the collective SNIa energy
input decreases, a subsonic outflow takes place, which gradually slows
until a central cooling catastrophe leads to the onset of an
inflow. At fixed $\lb$, any of the three phases wind, outflow or
inflow can be found at the present epoch, depending only on the
various depths and shapes of the potential well of the galaxies. In
this way both the large scatter in $\lx $ and the trend in the
spectral properties are accounted for at the same time: in the X-ray
bright galaxies the soft X-ray emitting gas dominates the emission,
being in the inflow phase, that resembles a cooling flow; in the X-ray
faint galaxies the hard stellar emission dominates, these being in the
wind phase.

Recent $ASCA$ observations seems to indicate that the SNIa's activity
is suppressed in early-type galaxies, which implies that the CDPR
scenario is essentially ruled out. Since this issue is far from closed
(\S 1.3), it is still worthwile to explore the effects of SNIa's on
the flows, using the updated rate given by recent optical surveys;
this is lower than that adopted by CDPR, which was 0.67 the standard
one estimated by Tammann (1982).  This aspect, together with the need
for changes in other galaxy properties crucial for the evolution of
hot gas flows, are discussed below.

   \begin{figure}[htbp]
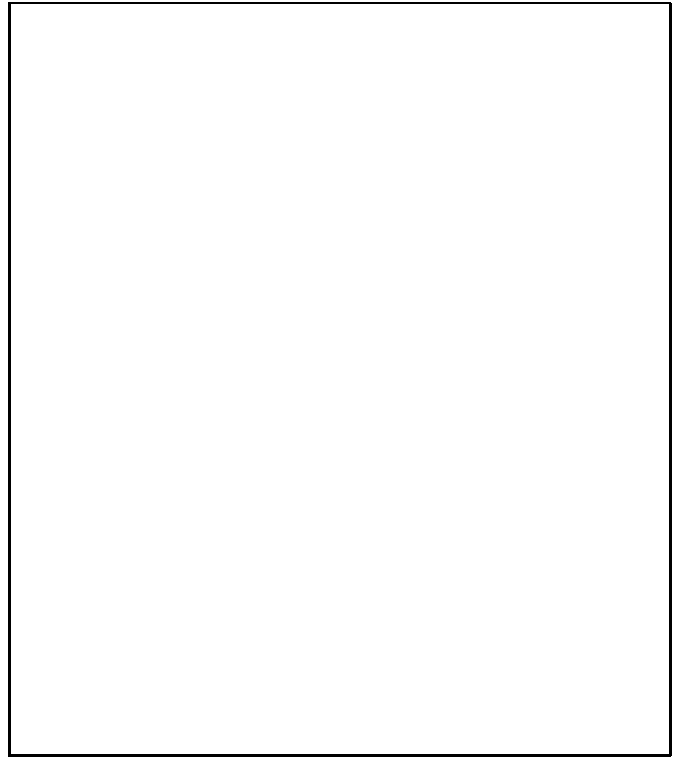

      \picplace{10cm}
      \caption{The $\lx-\lb$ diagram of early-type galaxies with X-ray emission
               detected by the {\it Einstein} satellite; X-ray fluxes are 
               from the final catalog of Fabbiano \etal (1992). Apparent B 
               magnitudes and distances are from Fabbiano et al. (1992) in 
               Fig.~1a, and from Donnelly et al. (1990) in Fig.~1b 
               (see Section 1). The dashed line gives an estimate for the 
               stellar source contribution to $\lx$ (from Kim \etal 1992).
               Also shown with triangles are the positions of the models 
               calculated here (see Section 3.5), to which the stellar source
               contribution has been added.}
      \label{Fig1}
   \end{figure}

\subsection {New stellar density profiles}

Previous numerical simulations of hot gas flows used the King (1972)
stellar density distribution mainly for computational ease. This
distribution has an inner region of constant density (the so called
``core'') of the order of a few hundred parsecs, which keeps the time
step of the numerical simulation reasonably small (a stellar density
increase implies a reduction of the characteristic hydrodynamical time
step). Another advantage of density profiles with a constant density
core is that the central regions can be resolved using a larger grid
size; this again allows a larger time step (see CDPR for a more
quantitative discussion). It is well known though that a much better
description of the surface brightness profiles of ellipticals is given
by the de Vaucouleurs (1948) law.  A very good fit to this law is
given by the Jaffe (1983) and by the Hernquist (1990) distributions,
that have the advantage that all their dynamical properties can be
expressed analytically. These distributions belong to the family of
the so called $\gamma$-models, that has been widely explored recently
(\cite{d93}; \cite{tremetal94}):

\begin{equation}
\rho (r)=M{(3-\gamma)\over 4\pi} {\rc\over r^{\gamma}(\rc+r)^{4-\gamma}}.
\end{equation}

The density profile of the Hernquist law has $\gamma=1$, while that of
the Jaffe law has $\gamma=2$.  Inside $\rc$ the density of
$\gamma$-models increases as $r^{-\gamma}$, a significant difference
with respect to King models: over a few hundreds of parsecs at the
center the $\gamma$-models are power laws. The existence of cores of
constant surface brightness has been definitively ruled out by ground
based observations (Lauer 1985, Kormendy 1985), and recently by the
Hubble Space Telescope (Jaffe et al. 1994; \cite{lauetal95}), that has
shown how the central surface brightness profile is described by a
power law as far in as can be observed, i.e., $\sim 10$ pc in
Virgo. From the point of view of a very accurate photometry, the
surface brightness law implied by eq. (1) cannot reproduce well both
the envelope and the very center of all the ellipticals
observed by HST. The Jaffe law, though, gives a general description of
ellipticals accurate enough for the treatment of hot gas flows, on the
scales that are relevant for the problem (from a few tens to several
thousands of parsecs).

\subsection {New dark matter estimates and distributions}

Attempts to estimate the amount of nonluminous mass in elliptical
galaxies have been made recently through extensive searches for
dynamical evidence of dark matter, either with systematic observations
of ionized gas probing the gravitational field (\cite{pizetal97}) or
with measurements of stellar velocity dispersions profiles out to
large radii (e.g., \cite{berteal94}; Carollo et al. 1995). These
optical studies are confined to within one or two effective radii
$\rref$, and typically find that luminous matter dominates the mass
distribution inside $\sim\rref$, while dark matter begins to be
dynamically important at 2--3 $\rref$. In particular, for a sample of
X-ray emitting galaxies, it has been found that dark matter halos are
not much more massive than the luminous component, with the
dark-to-luminous mass ratio $\mh/\ms\approx 1-6$; the value of
$\mh/\ms\sim 2$ is most common (\cite{saglieal93}). X-ray emission
from hot gas provides a great potential for mapping the mass of
ellipticals to larger distances (e.g., Fabian et al. 1986). The standard 
method employed derives
from the equation of hydrostatic equilibrium, and requires the
knowledge of the gas temperature profile. Attempts to apply this
technique to the {\it Einstein} data yielded a much larger component
of dark matter than found from optical data, but these results are
very uncertain because temperature profiles are poorly determined
(Forman, Jones, \& Tucker 1985; \cite{fab89}). Using superior X-ray
data provided by {\it ROSAT} and {\it ASCA}, Buote \& Canizares (1997)
proved that mass does not follow the optical light, out to many
$\rref$, in NGC720 and NGC1332. Adopting plausible gas and mass
models, they find $\mh/\ms >3$ for NGC1332, and $\mh/\ms >7$ in NGC720
at 90\% confidence; $\mh $ prevails exterior to $\rref$.  Similarly
Mushotzky \etal (1994) derive $\mh/\ms \sim 8$ within $8\rref$ for
NGC4636, and analogous results have been obtained from {\it ROSAT}
data of NGC507 and NGC499 by Kim \& Fabbiano (1995). {\it AXAF} will
have the combined spatial and spectral resolution to measure
accurately the presence of different spectral components, their
relative flux and their spatial distribution, and to produce more
accurate mass distribution from X-rays.

The radial density distribution of the dark haloes of ellipticals is
not well constrained by observations; theoretical arguments favor a
peaked profile (Ciotti \& Pellegrini 1992; Evans \& Collett 1997), and
high resolution numerical simulations of dissipationless collapse
produce a density distribution with $\gamma\simeq 1$ near the center
(\cite{dc91}; Navarro, Frenk, \& White 1996; \cite{w96}; Fukushige \&
Makino 1997, and references therein). Previous works studying the hot
gas flow evolution always used quasi isothermal haloes at least nine
times more massive than the luminous component. We are motivated now
to explore even the effects produced by dark haloes not as massive as
supposed before, and not quasi-isothermal.

\subsection{New Type Ia supernova rates}

Nearby SNIa rates in early-type galaxies have been carefully
reanalyzed recently, and this important ingredient of the simulations
of hot gas flows has been revised. From optical surveys it was
estimated to be 0.88$\,h^2$ SNu (\cite{tam82}), and then 0.98$\,h^2$
SNu (\cite{vdbt91}). Most recent estimates agree on lower values:
0.25$\,h^2$ SNu (\cite{vdbmc94}) and 0.24$\,h^2$ SNu (Turatto,
Cappellaro, \& Benetti 1994), where $h=H_{\circ}/100$ and 1 SNu = 1
SNIa per century per $10^{10}L_{\rm B\odot}$.

In principle, constraints on the SNIa rate can be given also by
estimates of the iron abundance in galactic flows (see, e.g.,
\cite{rcdp93}, and references therein). These were first attempted
using data from the {\it Ginga} satellite for NGC4472, NGC4636,
NGC1399 (Ohashi \etal 1990; Awaki \etal 1991; Ikebe \etal 1992), and then from
the {\it BBXRT} satellite for NGC1399, NGC4472 (\cite{serleal93}), and
more recently from {\it ASCA} with a superior spectral energy
resolution (\cite{loeweal94}; Awaki \etal 1994; Arimoto \etal 1997;
Matsumoto et al. 1997). Under the assumption of solar abundance ratio,
the analysis of all these data suggests a very low iron abundance,
consistent with no SNIa's enrichment and even lower than that of the
stellar component. However, some authors have found that more complex
multi-temperature models with higher abundance give a better fit of
the data (\cite{kimeal96}, Buote \& Fabian 1997).  Moreover, the
above results are based on iron line diagnostic tools whose
reliability has been questioned (e.g., \cite{arieal97}), especially
because of uncertainties affecting the Fe L-shell atomic physics for
temperatures less than 2 keV, that are typical of hot gas flows in
ellipticals (Liedahl, Osterheld, \& Goldstein 1995). Line emission
even in simple (e.g., isothermal) astrophysical plasmas needs to be
understood, and reliable fits to the data made, before secure
consequences concerning the SNIa rate can be drawn from X-ray
determined abundances.

\medskip

In summary, recent optical studies agree on a present epoch SNIa rate
much lower than previously used, and indicate that also the dark
matter content could be lower.  Moreover, cuspier density profiles,
especially for the stellar distribution, should be used. All this
raises the question of whether the CDPR scenario is altered by these
changes in the main ingredients of the problem, and more in general
what are the effects on the properties of hot gas flows. In this paper
we address these points, with a new set of hydrodynamical simulations.
In Section 2 we present the galaxy models, the source terms, and the
integration techniques. In Section 3 we discuss the main properties of
the evolution of gas flows in our new models, and we compare them with
the observations and the CDPR results. In Section 4 we discuss the
results using energetic arguments, and in Section 5 the main
conclusions are summarized.

\section{The models}

\subsection{Galaxy models}

Following the discussion in Sections 1.1 and 1.2, our model galaxies
are a superposition of a Jaffe density distribution for the luminous
matter $\rhos $ [eq. (1) with $M=\ms$, $r_{\rm c}=\rcs$, $\gamma=2$],
and of a Hernquist distribution for the dark matter [eq. (1) with
$M=\mh$, $r_{\rm c}= \rch$, $\gamma=1$]. We call these two-component
mass models JH models. Both density distributions have finite mass,
and so no truncation radius is applied.  The relation between the
scale length $\rcs$ and the effective radius for the Jaffe model is
$\rref\simeq 0.76\rcs$. We constrain the luminous body to lie on the
fundamental plane of elliptical galaxies (\cite{dd87}; Bender,
Burstein, \& Faber 1992).  The input observables are $\lb$, and the
central value of the projected velocity dispersion $\sgc$. The stellar
mass is related to $\sgc$ through the projected virial theorem, as
described in Ciotti, Lanzoni, \& Renzini (1996); the luminous matter
distribution is then completely defined, because it is assumed that
the stellar mass-to-light ratio is constant with radius, and the
orbital distribution is isotropic. The dark matter distribution is
determined by choosing the ratios of the dark to luminous mass
($\rap=\mh/\ms$), and of the dark to luminous core radius ($\beta
=\rch/\rcs$).  The basic dynamical properties of the JH models (radial
trend of the velocity dispersion, kinetic and potential energy, etc.)
are given in the Appendix.

   \begin{figure}[htbp]
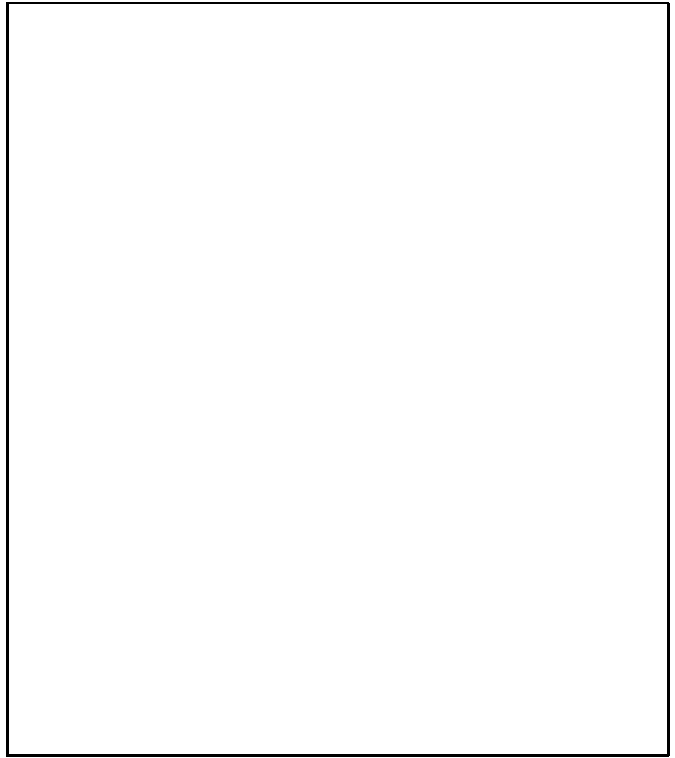

      \picplace{10cm}
      \caption{The cumulative mass profiles normalized to the total stellar 
               mass $\ms$ of JH models with $\rap =2$ and $\rap =9$; the cases
               $\beta=2$ and $\beta=4$ are shown in each panel. Within 
               $r=2\rref$, $\rap=0.6$ and $0.3$ respectively if $\beta=2$ and 
               $\beta=4$ (upper panel); $\rap=2.8$ and $1.1$ if $\beta=2$ and 
               $\beta=4$ (lower panel).}
      \label{Fig2}
   \end{figure} 

\subsection{Source terms}

The time evolving input ingredients of the numerical simulations are
the rates of stellar mass loss, of SNIa heating, and of thermalization
of the stellar velocity dispersion; these are calculated as in CDPR,
where a detailed description is given. Here we summarize only the main
properties of the input quantities. The stellar mass loss rate is
accurately described by ${\dot\ms}(t)\simeq
1.5\,\times\,10^{-11}\,\lb\, t_{15}^{-1.36}\; \msun\,{\rm yr}^{-1}$,
where $\lb$ is in $\lsun$, and $t_{15}$ is time in units of 15 Gyr; in
the numerical code the exact mass return prescribed by the stellar
evolution theory is used. The SNIa heating rate is parameterized as
$\lsn(t)=\Esn\,\Rsn(t)\,\lb = 7.1 \times 10^{30} \,\vt\,\lb\,
t_{15}^{-1.5}\;\es$, where $\Esn=10^{51}$ erg is the kinetic energy
injected in the ISM by one SNIa,
$\Rsn(t)=\vt\,0.88\,h^2\,t_{15}^{-1.5}$ SNu, and $h=0.5$. When $\vt=1$
and $t_{15}=1$ $\Rsn$ is the rate estimated by Tammann (1982).  $\lsn$
is assumed to be decreasing with time just faster than the stellar
mass loss rate. Unfortunately, given our enduring ignorance of the
nature of the SNIa's progenitors (e.g., Branch \etal 1995) the
evolution of the SNIa rate is not known; some arguments though, as an
observed pronounced correlation of the rate with the star formation
rate along the Hubble sequence, favour a declining rate (see also
Renzini \etal 1993; Ruiz-Lapuente, Burkert, \& Canal 1995; Renzini
1996). With the assumptions above the specific heating for the gas is
decreasing with time, and so we have the wind/outflow/inflow secular
evolution in the CDPR models ($\S$1). This direction of the evolution
of the flow is maintained as long as $\Rsn$ decreases faster than
${\dot\ms }$, and the results do not qualitatively depend on the
particular slope adopted for $\Rsn$; when this slope is much higher
than that of $\dot\ms$, though, too much iron is produced at early
times (see Renzini \etal 1993 for a more extensive discussion).

The heating given by the thermalization of the stellar velocity
dispersion at each radius is determined using the velocity dispersion
profile obtained by solving the Jeans equation in the global isotropic
case (see the Appendix).

\subsection {Numerical simulations}

The time-dependent equations of hydrodynamics with source terms, and
the numerical code used to solve them, are fully described in CDPR,
together with the initial conditions. We adopt here a higher spatial
resolution in the central regions of the galaxies: the central grid
spacing is 30 pc instead of 100 pc. This allows a better sampling of
the inner regions now described by a power law, and is adequate 
to highlight the differences in the flow properties between JH models
and King models plus a quasi isothermal dark halo.  The model galaxies
are initially devoid of gas, a situation produced by the galactic
winds established by type II supernovae, early in the evolution of
elliptical galaxies. The gas flow evolution is followed for 15 Gyrs.


   \begin{figure}[htbp]
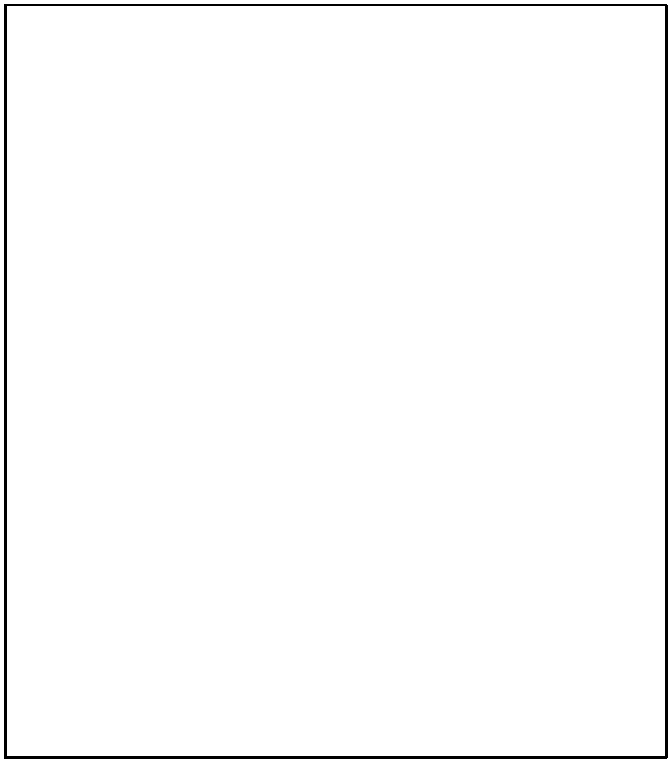

      \picplace{10cm}
      \caption{The evolution of the number density $n$, the temperature $T$ 
               and the velocity $v$ of the gas, in a typical PW case
               (the model with $\lb=10^{11}\lsun$, $\rap=2$, $\beta=4$ and 
               $\vt=0.25$ in Table 1). The solid line refers to $t=1.5$ Gyrs, 
               the dotted line to $t=4.5$ Gyrs, and the dashed line to $t=15$ 
               Gyrs. In this model the initially supersonic inflow slows 
               until it becomes fully subsonic at $t\simeq 6$ Gyrs; the outer 
               sonic radius starts at $\sim 2\rref$ and moves outward, until 
               it establishes at $\sim 8\rref$ from $t\simeq 10$ Gyrs on.}
      \label{Fig3}
   \end{figure} 

\section {Evolution of the gas flows}

We present here the results of the hydrodynamical simulations. The
basic input parameters and output quantities are shown in Table 1. Two
representative blue luminosities have been used to investigate the
typical gas flow behavior, $\lb=5\times 10^{10}\lsun$, and
$\lb=10^{11}\lsun$; the corresponding $\ms$ are $1.9\times
10^{11}\msun$ and $4.2\times 10^{11}\msun$. Various values of
$(\rap,\beta,\vt)$ are chosen.  $\vt=0.25$ is close to the most secure
current estimate coming from optical surveys (which is $\vt\sim 0.27$,
see $\S 1.3$); to consider the indications coming from the low iron
abundances given by X-ray data ($\S$1.2), we also explore the case
$\vt=0.1$; for comparison with the long-lived inflow case, we also use
$\vt =0$. The distribution of the dark mass is broader than that of
the luminous mass ($\beta=2$ or $\beta =4$); the dark mass prevails
outside one or two $\rref$ if $\rap=9$, or outside a few $\rref$ if
$\rap =2$ (Fig.~2).

A general feature of all the evolutionary sequences, except for those
with $\vt=0$, is to develop a {\it central} inflow at early times,
while the outer regions are still outgassing, i.e., a decoupled gas
flow (see Fig.~3). The inflow region is bordered by a stagnation
radius $\rst$ that may range from a small fraction of $\rref$ to many
$\rref$, at the end of the evolution (see Table 1). If $\rst$
maintains its position within a few $\rref$, we call the flow {\it
partial wind} (hereafter PW; see the representative case in Fig.~3 and
4); if it increases considerably to more than $10\,\rref$, we call the
resulting flow global inflow, even though this is strictly the case
only when $\vt=0$. In the PW case, radiative losses suppress an
outflow in the inner parts of the galaxy, causing a small inflow
region, but a wind can be sustained in the external parts, where the
gas density is much lower and the gas is also less tightly bound.  The
X-ray luminosity of PWs and inflows steadly decreases with time,
following the decrease of $\dot\ms$ and of the heating.  Note that the
stagnation points resulting from the simulations are well outside the
first grid point, so the flow is properly resolved.

Another general trend in the results is that, at fixed $\rap$, as
$\vt$ and/or $\beta$ decrease, the larger are the inflow
region, $\lx$, the mass accreted to the center $\min$, the mass flowing
to the center per unit time $\dot\min$, and the smaller is $\dot\mou$,
the mass escaping the galaxy per unit time, because the heating is
lower and/or the dark mass concentration is higher.

   \begin{figure}[htbp]
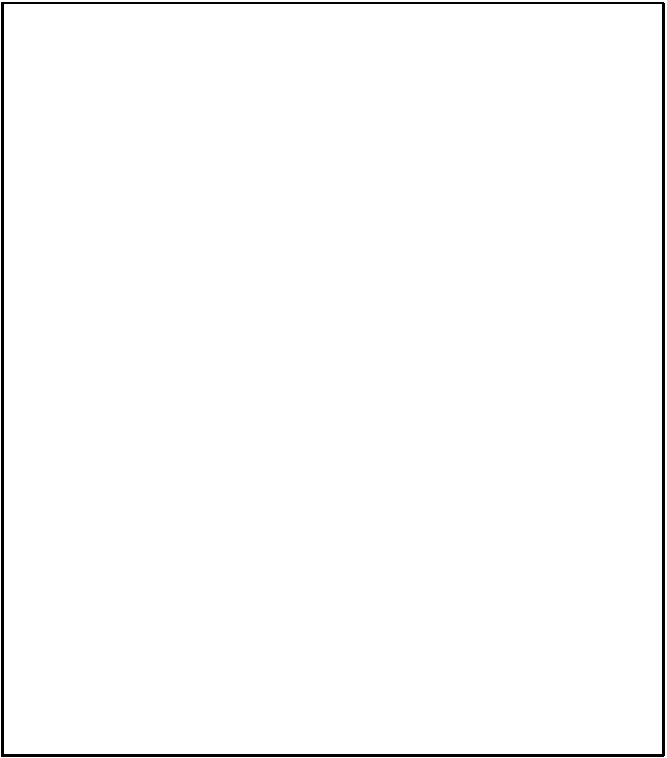

      \picplace{10cm}
      \caption{Upper panel: the time evolution of the mass loss $\dot \ms$, of 
               the mass inflowing into the central sink $\dot\min$, and of the
               mass lost by the galaxy $\dot \mou$. Lower panel: the time 
               evolution of the stagnation radius $\rst$. All refers to the 
               representative model chosen for Fig.~3.}
      \label{Fig4}
   \end{figure} 

\subsection {$\lb=5\times 10^{10}\lsun$ models}

For $\rap=2$, and $\vt >0$, these galaxies develop PWs with $\rst$
varying from $0.02\,\rref$ to $\simeq 3\,\rref$. When $\vt >0.25$ a
global wind persists until the present time, and so $\lx$ is very
low. Even when $\vt=0.25$ the hot gas, in PW, shows such a low $\lx$
that the observed X-ray emission must be largely produced by the
stellar sources, whose emission has been estimated to be of the order
of $\sim 10^{40}\;\es$ (see Fig.~1).  When $\vt =0.1$, the final
$\lx\,\lsim\, 5\times 10^{40}\;\es$, and the hot gas starts to
dominate the X-ray emission of the galaxy. Significantly higher $\lx$
values are obtained by increasing the gas temperature, not just by
retaining more gas; in fact the model with $\rap=2$ and $\vt=0$,
 a global inflow that lasts for the whole galaxy lifetime, has just
$\lx\sim 7\times 10^{40}\;\es$.  To increase the gas
temperature {\it and} $\lx$, the dark matter content must be increased
first, because global winds are obtained by just increasing the SNIa
heating. A deeper potential well by itself produces hotter gas,
because of a higher gravitational compression during the inflow;
moreover, it can be coupled to a larger inflow region, if $\vt$ is
kept constant, or to a higher heating without full degassing if $\vt $
is increased.

For $\rap=9$ and $\vt\lsim 0.25$ the gas is always in a global inflow
at the present time. The highest $\lx\sim 2\times 10^{41}\;\es$ is
obtained when $\vt=0.25$ and $\beta=2$, i.e., when the heating by
SNIa's and gravitational compression is the highest possible for the
chosen ranges of these parameters. This $\lx$ is still lower
than the highest $\lx$ observed at $\lb=5\times 10^{10} \lsun$
(Fig.~1), a problem that was known also from previous works (see the
references cited in the Introduction). However, $\lx$ as high as $\sim
10^{42}\;\es$ are comparable to those typical of poor clusters rather
than single galaxies; at least some of them could be explained with
accretion effects (see, e.g., \cite{rcdp93}; \cite{bt95};
\cite{kf95}).

   \begin{table*}
      \caption{Results of the JH model evolution. $\lx$ is calculated for the 
              (0.2--4) keV band; $T$ is the emission weighted temperature of 
               the flow. $\min$ is the gas mass flown to the center in 15 
               Gyrs; $\dot\min$ and $\dot\mou$ are the gas masses that flow 
               to the central sink and are lost from the galaxy, per unit time,
               at 15 Gyrs. At this time $\dot\ms=0.72$ $\msun$/yr, for the 
               model with $\lb=5\times 10^{10}\lsun$, and 
               $\dot\ms=1.44\msun$/yr for the model with $\lb= 10^{11}\lsun$. 
               $\chi$ is defined in $\S 4$. }
         \label{Table1}
      \picplace{10cm}
   \end{table*}

Note that for PWs ($\rap=2$) $\lx$ is higher when $\vt$ is lower,
because the inflow region is larger. For global inflows ($\rap =9$)
$\lx$ is higher when $\vt$ is also higher, because the gas is
hotter. Altogether, the variation in $\lx$ in Table 1 is a factor
of $\sim 100$; it is of a factor of $\sim 30$ when $\rap=2$, and just
$\sim 2$ when $\rap =9$.

\subsection {$\lb=10^{11}\lsun$ models}

Models with $\rap=2$ again develop PWs, with quite larger central
inflow regions than above for the same range of $\vt$ and $\beta$
values, due to a deeper potential well. The latter makes also the gas
hotter. Larger inflow regions and higher gas temperatures, coupled to
a larger mass return rate, make the X-ray emission in these galaxies
always higher than that of the models described in $\S 3.1$, for the
same $(\rap,\beta,\vt)$. $\lx$ goes from $3\times 10^{40}\;\es$ to
$2\times 10^{41}\;\es$, when the flow is a PW. So, it is always
comparable to or higher than the estimate of the stellar X-ray
emission at $\lb=10^{11}\lsun$ (Fig.~1); this is in agreement with the
fact that in Fig.~1 all galaxies lie above this estimate, for
$\lb\gsim 10^{11} \lsun$.

As expected, we always have global inflows if $\rap =9$, and the
highest $\lx$ is reached when $\rap=9$ and $\vt=0.25$, i.e., again the
most efficient way of increasing $\lx$ when the flow is not decoupled
is to increase $\rap$, and then $\vt$.  The spread in $\lx$ is of a
factor of $\sim 10$ if $\rap=2$, and very small for global
inflows. The total spread is roughly of a factor of 30, lower than for
$\lb=5\times 10^{10}\lsun$.

   \begin{table*}
      \caption{Results of additional JH model evolutions.
               All quantities are defined as in Table 1. After 15 Gyrs $\dot
               \ms=0.32$ $\msun$/yr, for the model with $\lb=2.2\times 
               10^{10}\lsun$, and $\dot\ms=1.02$ $\msun$/yr for the model with
               $\lb=7.1\times 10^{10}\lsun$. At fixed $\lb$, the models with 
               the larger $\rref$ are  less concentrated than required for 
               them to lay exactly on the fundamental plane. }
      \label{Table2}
      \picplace{10cm}
   \end{table*}

\subsection {Other models}

In order to investigate how common is the PW phase, we have run two
more series of evolutionary sequences corresponding to a low
$\lb=2.2\times 10^{10}\lsun$, and to $\lb= 7.1\times 10^{10}\lsun$;
moreover, also the case $\rap =4$ has been explored. The results are
summarized in Table 2. PWs populate again most of the parameter
space. The X-ray luminosity of $\lb=2.2\times 10^{10}\lsun$ models
remains at low values even at high $\rap$, in agreement with the
observations (Fig.~1); for many of these galaxies $\lx$ is likely to
be dominated by the stellar emission.

\subsection {Emission temperatures and X-ray surface brightness profiles}

Emission temperatures of the hot gas have been calculated recently by
Matsumoto et al. (1997) and Buote \& Fabian (1997), using $ASCA$ data
and two-temperature model fitting, for about 20 early-type galaxies.
These temperatures cover the range $\sim 0.3-1.0$ keV, and the bulk of values lies
within $0.5-0.8$ keV.  The X-ray luminosity weighted emission
temperatures of the models are given in Tables 1 and 2. They lie in
the range $\sim 0.3-0.8$ keV, and so are in good agreement with those
observed. They increase with central velocity dispersion, dark
matter mass, and SNIa rate. Note how a substantial SNIa's heating
helps obtaining temperatures close to those observed; other possible
sources of heating not studied here are the presence of an external
pressure due to an intracluster or intragroup medium (e.g., Bertin \&
Toniazzo 1995), and that of a central black hole (Ciotti \& Ostriker
1997).

The comparison with the observed X-ray surface brightness profiles
$\SX (R)$ is more delicate, because there is not a typical profile,
nor a range of typical profiles. In the next, first we examine what
kind of $\SX (R)$ are shown by PWs (which are the resulting flow
regimes of the bulk of the galaxies, in our scenario), and then we
discuss how $\SX (R)$ of our models compare with the
available observations.

In Fig.~5 we compare the {\it
shapes} of the X-ray profiles of partial wind solutions with different
stagnation radii with that of a global inflow with $\vt =0$ (which
closely resembles the cooling flow solution). When the stagnation
radius is larger than the optical effective radius, the shape of $\SX
(R)$ is indistinguishable from that of the inflow, over most of the
galaxy. PWs with very small stagnation radii are instead less steep
outside $\sim\rref$; this is because in the external regions the gas density
is actually lower for the PW than for the inflow, but the prevailing
effect is given by the higher gas temperature of the PW, produced by
the higher $\vt$.  So, judging from the shape of the $\SX (R)$
profile, inflows and PWs should be indistinguishable by X-ray
observations, if $\rst$ is a few $\rref$.  Of course, when $\rst$ is
very small, $\lx$ is very low too, and so these PWs are much less
detectable by X-ray observations than global inflows.

Now let's turn to the comparison with the observations.  When $L_X$ of
the gas flow is very low, and the X-ray emission is dominated by the
contribution of the stellar sources, our models exhibit the X-ray
profile of the underlying stellar population. No detailed $\SX (R)$
for low $\lx/\lb$ galaxies is available yet from the observations.  When $L_X$
of the flow is high, the observations -- especially those produced by
$ROSAT$ -- show a large variety of shapes for $\SX (R)$, often
modified also by the interaction with the environment (e.g.,
Trinchieri, Fabbiano \& Kim 1997). Using {\it Einstein} data,
Trinchieri, Fabbiano, \& Canizares (1986) found that the observed $\SX
(R)$ tend to follow the optical ones, for few best studied X-ray
bright galaxies. In our scenario such galaxies host PWs with large
stagnation radii, or inflows; these flow phases have the same problems
as cooling flows to reproduce X-ray profiles following the
optical ones: they are too peaked (see Fig.~5). They can be brought in
agreement with the observed $\SX (R)$ by the introduction of
distributed mass deposition due to thermal instabilities in the hot
gas, as shown and discussed in detail by \cite{sa89}, and
\cite{bt95}; another possibility is stationary convective accretion
onto a central massive black hole (\cite{tb93}), or unstable accretion
(Ciotti \& Ostriker 1997).

A detailed comparison with $\SX (R)$ observed for a specific galaxy
requires a large number of simulations to find whether there is a
combination of input parameters (with the possible addition of the
effect of the environment, and/or of a central black hole) that gives
a model reproducing the observations. This goes beyond the scope of
this paper; it has been done successfully for NGC4365, in the framework of the
CDPR scenario, by Pellegrini \& Fabbiano (1994).

   \begin{figure}[htbp]
      \picplace{10cm}
      \caption{The X-ray surface brightness profile of three models with 
               $\lb=5\times 10^{10}\lsun$, $\rap =2$ and $\beta=2$: the solid 
               line corresponds to a global inflow ($\vt=0$), the dotted line 
               to a PW with $\rst=3.5\rref$ (the model with $\vt=0.1$ in Table
               1), and the dashed line to a PW with $\rst=0.04\rref$ (the 
               model with $\vt=0.25$ in Table 1). The heavy dot-dashed line is
               the optical surface brightness profile in arbitrary units.}
      \label{Fig5}
   \end{figure} 

\subsection {Effects of the new ingredients}

We summarize here the main properties of the flows, and the main
differences with the results obtained by CDPR, produced by the new
mass distributions and by the reduction of $\vt$ and $\rap$. The
global wind phase disappears for all galaxies except the smallest ones
(see the case $\lb =2.2\times 10^{10}\lsun$ in Table 2): a central
inflow, even though very small, is always present from the beginning
of the evolution. As a consequence the outflow phase -- the transition
from a global wind to a global inflow in CDPR -- also
disappears. Large $\lx$ variations are produced in the JH models by
the different size of the central inflow region. Most of the observed
spread in $\lx$ at fixed $\lb$ can be reproduced again, as it was in
CDPR (Fig.~1): besides the observed variation from galaxy to galaxy in
$\sgc$ and in the concentration $\rref$, a spread in the dark matter
content, and/or a variation in the SNIa rate can produce large
variations in $\rst$ and in the gas temperature, and then in $\lx$ (of
even a factor of $\sim 100$, Table 1 and Fig.~1). If $\rap$ remains at
low values ($\rap\lsim 4$), though, it is difficult to reproduce $\lx$
higher than a few times $10^{41} \;\es$ (Table 2), and so to cover all
the observed $\lx$ variation. Moreover, for a given range of variation
of $\rap$ and $\vt$, more luminous galaxies show less scatter in
$\lx$.

Another property of the flows in JH models is to accumulate cold gas
at their centers, during the evolution ($\min $ in Tables 1 and
2). This mass when $\vt >0$ ranges from $10^9\;\msun $ to $2\times
10^{11}\;\msun$. Evidence of cold gas at the center of bright X-ray
galaxies has come recently from the high column densities ($N_H$)
required to obtain good fits of their X-ray spectra. The {\it BBXRT}
data require $\sim 10^{10}\;\msun$ of cold gas in NGC1399
(\cite{serleal93}); to fit the {\it ASCA} data of NGC4472, $3\times
10^{10}\; \msun$ of cold material are needed (\cite{aweal94}); {\it
ASCA} data require central column densities larger than the Galactic
value also for NGC4636, NGC4406, NGC720, NGC1399, NGC1404, NGC4374
(\cite{arieal97}).  It has been suggested to explain this finding as
the accumulation of cold material by a steady state cooling flow.
This accumulation could have been produced by a PW as well; moreover,
even different amounts of cold material can be naturally explained in
this scenario (see the large variations of $\min$ in Table 1). So, the
finding of cold matter at the center of early-type galaxies cannot be
considered as evidence for a long lived cooling flow.  Note however
that the accuracy of these $N_H$ absorption measurements could be
affected by the uncertainties plaguing plasma spectra, and producing
the iron-L problem (\S 1.3).

\section{An energetic explanation for the occurrence of the PW in 
         different galaxy models}

The origin of PWs, and of the flows of a different nature obtained by
CDPR, can be explained qualitatively in terms of the {\it local}
$\chil$ function. The {\it global} $\chi$ is defined as the ratio
between the energy required to steadily extract the gas lost by the
stars per unit time from the galactic potential well ($\lgm$), and the
heating supplied by the SNIa explosions plus the thermalization of the
stellar velocity dispersion ($\lsi$). $\chi$ was introduced by CDPR to
predict the expected flow phases in King galaxies, and was shown to
work remarkably well by the numerical simulations: when $\chi >1$ the
flow turned out to be an inflow, when $\chi <1$ a wind. In Tables 1
and 2 most of the resulting flow phases are not global, but PWs, and
so we are moved to investigate in more detail the distribution of the
energetics inside the galaxy models. This distribution in fact can be
much different for two galaxies with the same global $\chi$. The local
$\chil$ is defined as

\begin{equation}
\chil(r)\equiv {d\lgm(r)\over d\lsi(r)+d\lsn(r)},
\end{equation}
where
\begin{equation}
d\lgm(r)=4\pi\alpha(t)\rhos(r)|\phit(r)|r^2dr, 
\end{equation}
\begin{equation}
d\lsi(r)=6\pi\alpha(t)\rhos(r)\ss^2(r)r^2dr, 
\end{equation}
and
\begin{equation}
d\lsn(r)={\lsn (t)\over\ms}4\pi\rhos(r) r^2dr, 
\end{equation}
where $\phit=\phis+\phih$ is the total potential, $\ss $ is the
one-dimensional isotropic velocity dispersion obtained by solving the
Jeans equations for the adopted mass model (e.g., \cite{bt87}), and
$\alpha(t)=\dot\ms/\ms$ is the specific mass return rate. The global
$\chi$ is obtained by replacing the differentials in equation (2) with
their integrals over the whole galaxy.

The differences in the flow behavior of JH models and models with a
central constant density region can be understood with the aid of
Fig.~6, where $\chil$ is shown for four representative one-component
models. These have the same global $\chi$, and so from an energetical
point of view are globally equivalent. Three of them are
$\gamma$-models, with $\gamma=2,1,0$, i.e., they are the Jaffe model,
the Hernquist model, and a model with a core. The fourth density
distribution is described in the Appendix, and is flatter at the
center than the $\gamma=0$ model, but for large $r$ it decreases
$\propto r^{-4}$, as all the $\gamma$-models. This distribution is
studied for its similarity with the King (1972) model, that needs the
annoying introduction of a truncation radius, and has more complicated
dynamical properties.

   \begin{figure}[htbp]
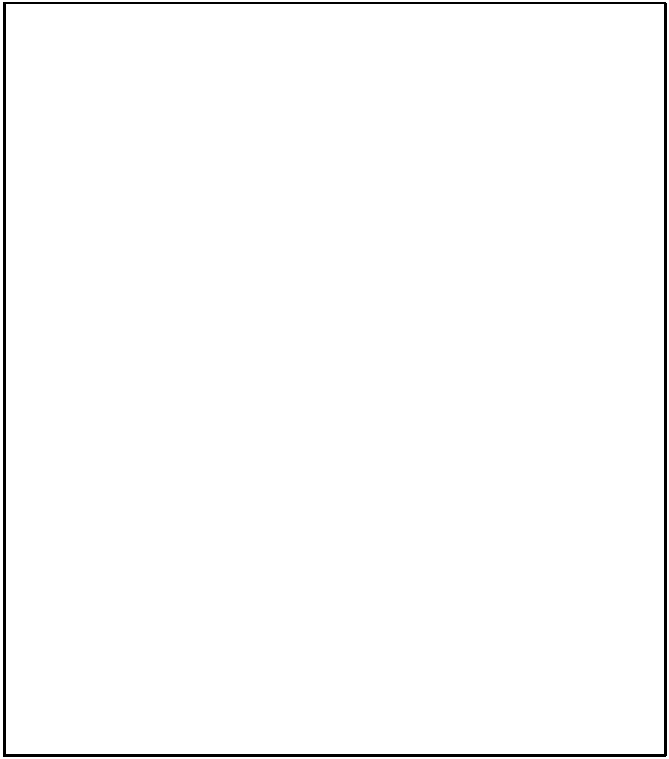

      \picplace{10cm}
      \caption{The radial trend of $\chil$ as a function of $r/\rref$ for the
               models considered in $\S 4$. The ratio $\rref/\rcs$ is given by 
               0.76, 1.82, 2.87, 1.73 respectively for the Jaffe (solid line),
               Hernquist (dotted), $\gamma=0$ (short-dashed) and the model 
               described in the Appendix (long-dashed).}
      \label{Fig6}
   \end{figure}

It is apparent from Fig.~6 that the steeper the density profile, the
stronger is the variation of $\chil$ across the galaxy, and the higher
are the values that it reaches at the center. This explains why a
strong decoupling can be present in the flow of highly concentrated
systems: a $\chil$ significantly higher than unity in the central regions
produces a central inflow, while in the external parts a degassing is
energetically possible because $\chil<1$. {\it This can happen when
the global $\chi$ is either $>1$ or $<1$} (see Fig.~6 and Table 1), so
this parameter is not a good indicator of the flow phase for highly
concentrated systems. If $\chi$ decreases, the radius where $\chil=1$
moves inward for all the models, and so a larger part of the galaxy is
degassing.  Note a fundamental difference in the trend of $\chil (r) $
as a function of $\chi$ for models with a core and cuspier models:
while in cuspy models there is always a region with $\chil >1$, even
though small, this is not the case for core models, in which such a
region suddenly appears by increasing $\chi$, with a size larger than
the core.

A second effect on $\chil$ is produced by the dark halo. If this halo
is more diffuse than the stellar mass, increasing $\rap$ makes more
bound the external regions, when $\chi$ is kept constant, and so the
radius at which $\chil=1$ moves outward (see Fig.~7 for the case of JH
models).

The trend of $\chil (r)$ for models with a core explains the results
of the numerical simulations for King models plus diffuse
quasi-isothermal dark halos obtained by CDPR and Pellegrini \&
Fabbiano (1994).  The flow phases found by CDPR with $\rap =9$ were
all global; with low $\rap$, instead, the flow can be decoupled again,
as found by Pellegrini \& Fabbiano (1994) in their detailed modeling
of the X-ray properties of two ellipticals. This because with high
$\rap$ the region with $\chil\approx$ constant is very large (of the
order of the core radius of the dark halo), while with low $\rap$ this
region is of the order of the stellar core radius.


In summary, the combined effect of SNIa's and dark matter is more
varied for models with a core than for JH models: for a large range of
$\rap$ values, the latter keep in PWs with a varying $\rst$, while the
former can be in wind, PWs, outflows or inflows.

   \begin{figure}[htbp]
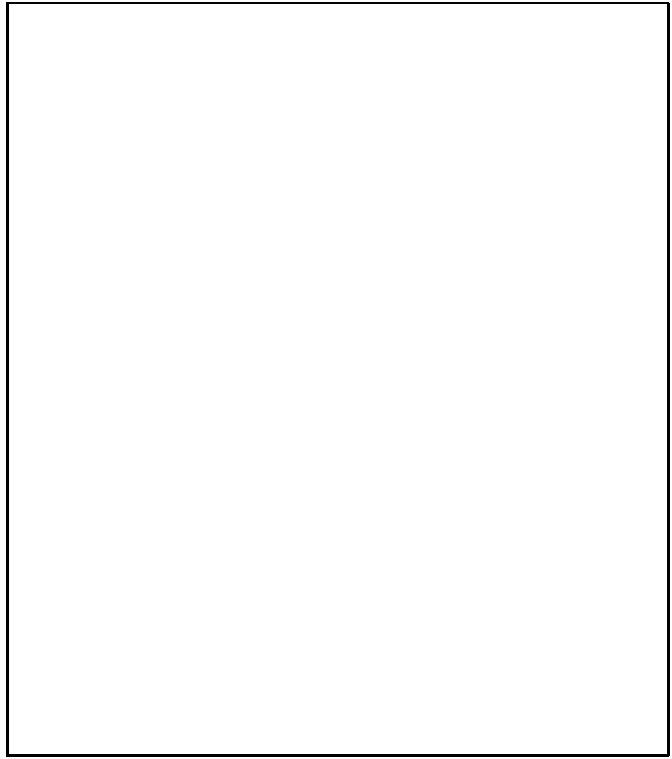

      \picplace{10cm}
      \caption{The radial trend of $\chil$ as a function of $r/\rref$ for the
               JH model. The different curves are labelled with the value of
               $\rap$, while the global $\chi $ and $\beta$ are constant at
               the values indicated.}
      \label{Fig7}
   \end{figure} 

\section{Conclusions}

Using evolutionary hydrodynamical simulations, we have investigated
the properties of hot gas flows in elliptical galaxies described by a
Jaffe stellar density profile, and containing various amounts of dark
matter, distributed as a Hernquist law; the rate of SNIa explosions is
one-fourth of the Tammann's rate, or lower.  The results can be
summarized as follows.

\smallskip

The simulations show that most of the gas flows are strongly
decoupled, i.e., they develop an inflow in the central region of the
galaxy, while the external parts are still degassing\footnote{Note
that decoupled flows are present also in flat and rotating models with
low $\vt$, albeit for reasons different from those causing the PWs
studied here (D'Ercole \& Ciotti 1997).}.  The stagnation radius of
these flows (PWs) may range from a small fraction of the optical
effective radius to many $\rref$.  This situation is present even when
the global energy balance of the flow indicates that the energy inputs
from SNIa's and from the thermalization of the stellar random motions
is high enough to unbind all the gas ($\chi <1$). Alternatively, the gas
can be outflowing from the outer part of the galaxy even when the
global energy balance indicates that the energy available is less than
needed to unbind it all ($\chi >1$). So, $\chi$ is not a good
indicator of the flow phase; larger $\chi $ values, though, correspond
to larger stagnation radii, for a fixed $\lb$.  Global inflows are
produced when the dark matter content is so high to bind the gas over
the whole galaxy.

The X-ray luminosity of PWs is higher when $\vt$ is lower, because the
inflow region is larger; for global inflows $\lx$ is higher when $\vt$
is also higher, because the gas is hotter.  The key factor causing
large $\lx$ variations in JH models is the size of the central inflow
region.  The lowest $\lx/\lb$ values observed can be easily reproduced
by PWs; high dark matter contents are required to approach the highest
$\lx$ observed.

The shape of the X-ray surface brightness profile of PWs is close to
that of global inflows, over most of the optical image, if the
stagnation radius is a few $\rref$; it is externally less peaked than
that, if $\rst$ is a fraction of $\rref$, i.e., when the gas is hotter
than in the inflow case.

\smallskip 

The strong decoupling of the flow is explained by the radial trend of
the local energy balance $\chil$, that is very peaked for steep mass
density profiles as in the JH models.  Previously used density
profiles which were flat at the center -- such as the King models plus
quasi-isothermal dark halos -- showed a larger tendency to have global
flow phases, because also their $\chil (r)$ is flatter.  So, the
change of the mass profile has an important effect: both numerical
simulations and analytical calculations show how peaked density
profiles preferentially develop decoupled gas flows (in the sense that
the region of the parameter space populated by PWs is large compared
to that of inflows or winds).

\smallskip 

The results of our simulations compare with the observations in two
main aspects. The first is that most of the observed spread in $\lx$
shown by the data is easily reproduced: the highest $\lx$ are again,
as in CDPR, associated with global inflows, while the bulk of the
galaxies are now in PW, rather than in outflow; global winds are
present only in the smallest galaxies.

The second is that the presence of cold gas at the center of Es, and a
peaked X-ray surface brightness profile, cannot be unequivocally
associated to a cooling flow: a PW could be present as well, with a
significant part of the galaxy degassing.  Particularly, a galaxy can
have $\vt=0.25$ and still a non negligible amount of cold matter at
the center. In addition, since $\rst$ varies over a wide range of radii, PWs 
can accumulate largely varying quantities of cold gas.

\begin{acknowledgements}
We wish to thank James Binney, Annibale D'Ercole, Paul Goudfrooij and
Alvio Renzini for stimulating discussions.  This work has been
partially supported by the contract ASI-95-RS-152.
\end{acknowledgements}

\section{Appendix}

\subsection{The JH models}

We derive here the main properties of the JH models that have been
used in the numerical simulations and in the computations of $\chi$
and $\chil$. The reported quantities are dimensionless, and the
normalization constants are expressed in terms of the total stellar
mass $\ms$, the core radius $\rcs$ and the gravitational constant $G$;
again $\rap=\mh/\ms$ and $\beta=\rch/\rcs$.  The cumulative mass
inside $s=r/\rcs$ and the total potential are given by:
\begin{equation}
\tilde M(s)={s\over 1+s}+{\rap s^2\over (\beta+s)^2},
\end{equation}
\begin{equation}
\tilde\phit (s) =-\ln\left({1+s\over s}\right)-{\rap\over\beta+s}.
\end{equation}
In the hydrodynamical simulations we used the globally isotropic
stellar velocity dispersion profile, that we calculate here. For any
two-component model the velocity dispersion profile is given by $\rhos
\ss^2=\iss (s)+ \rap \ish (s)$, where $\ss$ is the 1-dimensional radial
velocity dispersion.  The first term is the isotropic velocity
dispersion for the Jaffe model:
\begin{equation}
\tilde\iss (s) ={(6s^2+6s-1)(2s+1)\over 8\pi s^2(1+s)^2}+
                  {3\over 2\pi}\ln{1+s\over s},
\end{equation} 
and the second one describes the contribution due to the Hernquist
dark halo:
\begin{eqnarray}
\tilde\ish (s) ={\ln(s)(\beta+1)\over 2\pi\beta^3}
         -{\ln(1+s)(\beta-2)\over 2\pi(\beta-1)^3}-
         {\ln(\beta+s)(2\beta-1)\over 2\pi\beta^3(\beta-1)^3}+\nonumber \\
{2(\beta^2-\beta+1)s^2+(\beta+1)(2\beta^2-3\beta+2)s+\beta(\beta-1)^2\over
  4\pi\beta^2(\beta+s)(1+s)}.
\end{eqnarray} 
Physically acceptable limits can be easily found for $\beta=0$ (thus
mimicking a black hole at the center of a Jaffe model) and for
$\beta=1$.

The cumulative potential and kinetic energies inside $r$ can be found
analytically for this two-component model, but for simplicity we
report here only their total values. The total potential energy is
given by $U_*= \uss +\rap\ush$, where
\begin{equation}
\tilde\uss=-{1\over 2},
\end{equation}
and 
\begin{equation}
\tilde\ush =-{\beta-1-\ln(\beta)\over (\beta -1)^2}.
\end{equation}
By integrating over the whole galaxy equation (3), we have
$\lgm=\alpha (t) (G\ms^2/\rcs)(2|\tilde\uss |+\rap |\tilde\ush |)$.
The kinetic energy is again the sum of two different contributions,
$K_*= \kss+\rap\ksh$, where
\begin{equation}
\tilde\kss ={1\over 4},
\end{equation}
and
\begin{equation}
\tilde\ksh={(\beta+1)\ln(\beta)+2(1-\beta)\over 2(\beta-1)^3}.
\end{equation}
The integration over the whole galaxy of equation (4) gives $\lsi
=\alpha (t) (G\ms^2/\rcs)(\tilde\kss +\rap\tilde\ksh )$.

\subsection{A centrally flat model}

Here we derive the dynamical quantities corresponding to the density
distribution
\begin{equation}
\tilde\rhos(r)={1\over\pi^2}{1\over(1+s^2)^2}.
\end{equation}
The mass contained inside $r$ is given by
\begin{equation}
\tilde\ms(s)={2\over\pi}\left[\arctan(s)-{s\over 1+s^2}\right],
\end{equation}
and the potential is
\begin{equation}
\tilde\phi_*(s)=-{2\over\pi}{\arctan(s)\over s}.
\end{equation}
The globally isotropic velocity dispersion is given by:
\begin{eqnarray}
\tilde\iss (s)={3\arctan(s)^2\over 2 \pi^3 }+
               {(2+3s^2)\arctan(s)\over s(1+s^2)\pi^3 }+
               {4+3s^2\over 2(1+s^2)^2 \pi^3 } - {3\over 8 \pi} .
\end{eqnarray}
As for the JH model, the potential and the kinetic energy inside the
radius $r$ can be expressed analytically, but here we give only their
total values, being the only ingredient required in the definition of
$\chi$:
\begin{equation}
\tilde\uss=-{1\over 2\pi},
\end{equation}
and
\begin{equation}
\tilde \kss={1\over 4\pi}.
\end{equation}
Finally, the projected distribution associated to $\tilde\rhos$ is:
\begin{equation}
\tilde\Sigma_* (R)={1\over 2\pi (1+\tilde R^2)^{3/2}},
\end{equation}
from which one obtains $\rref=\sqrt{3}\rcs$.

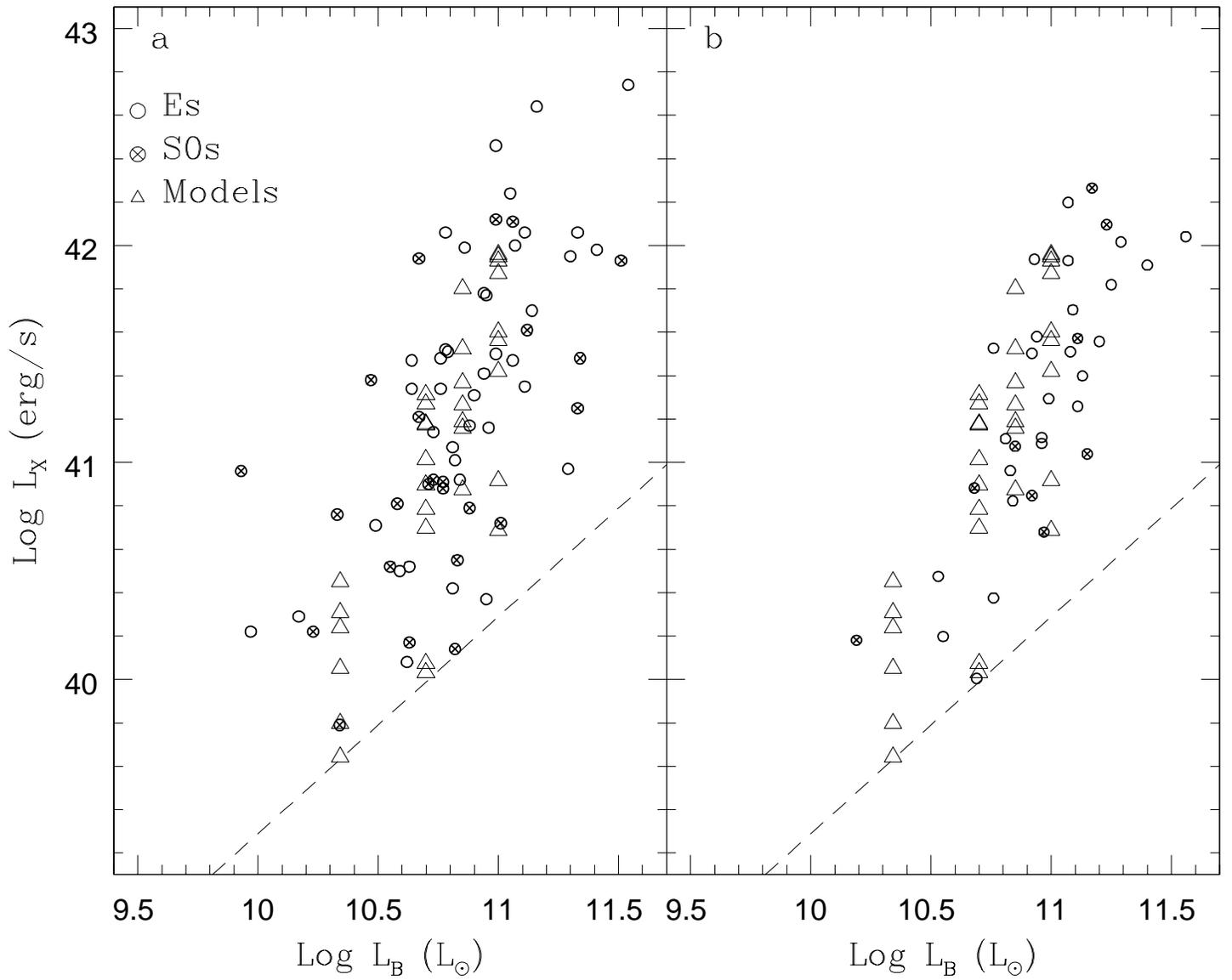
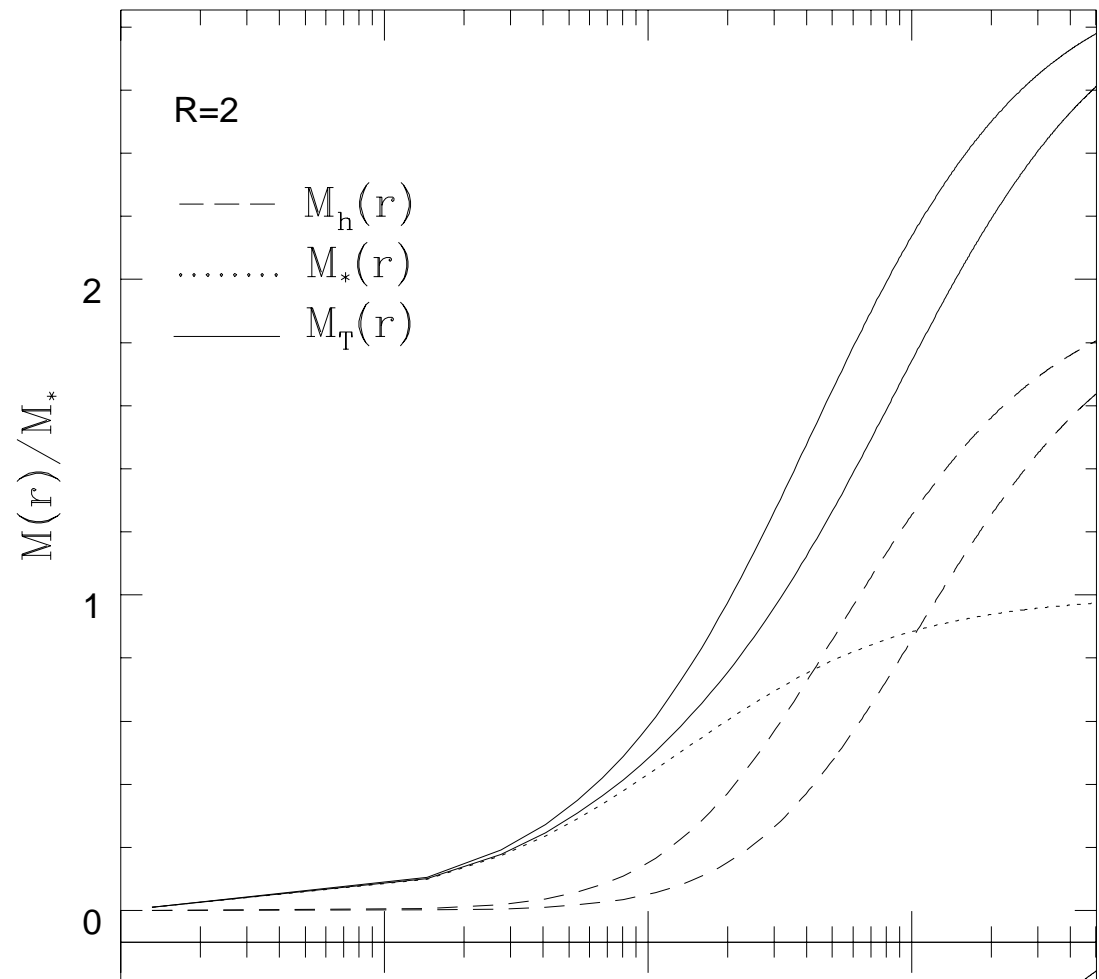
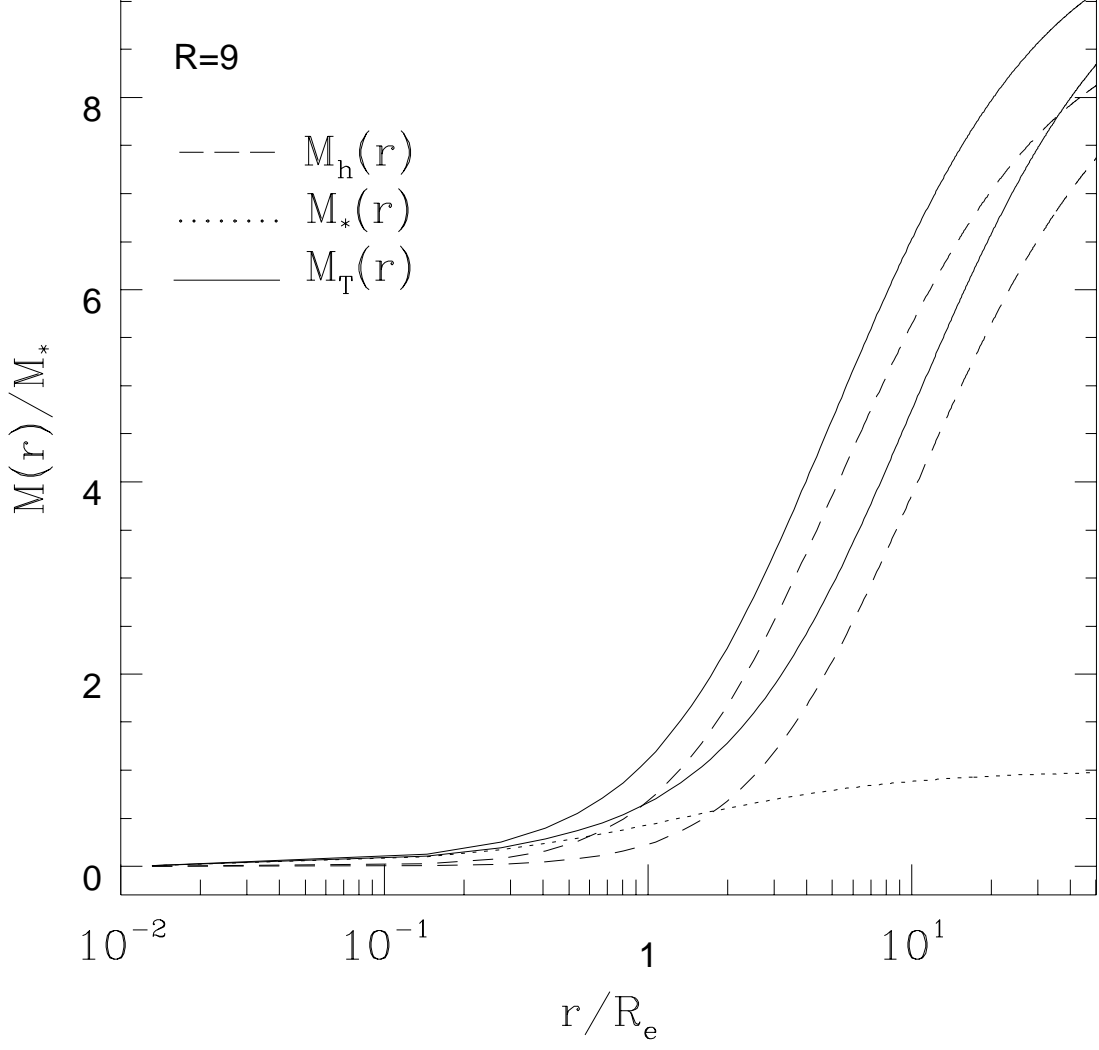
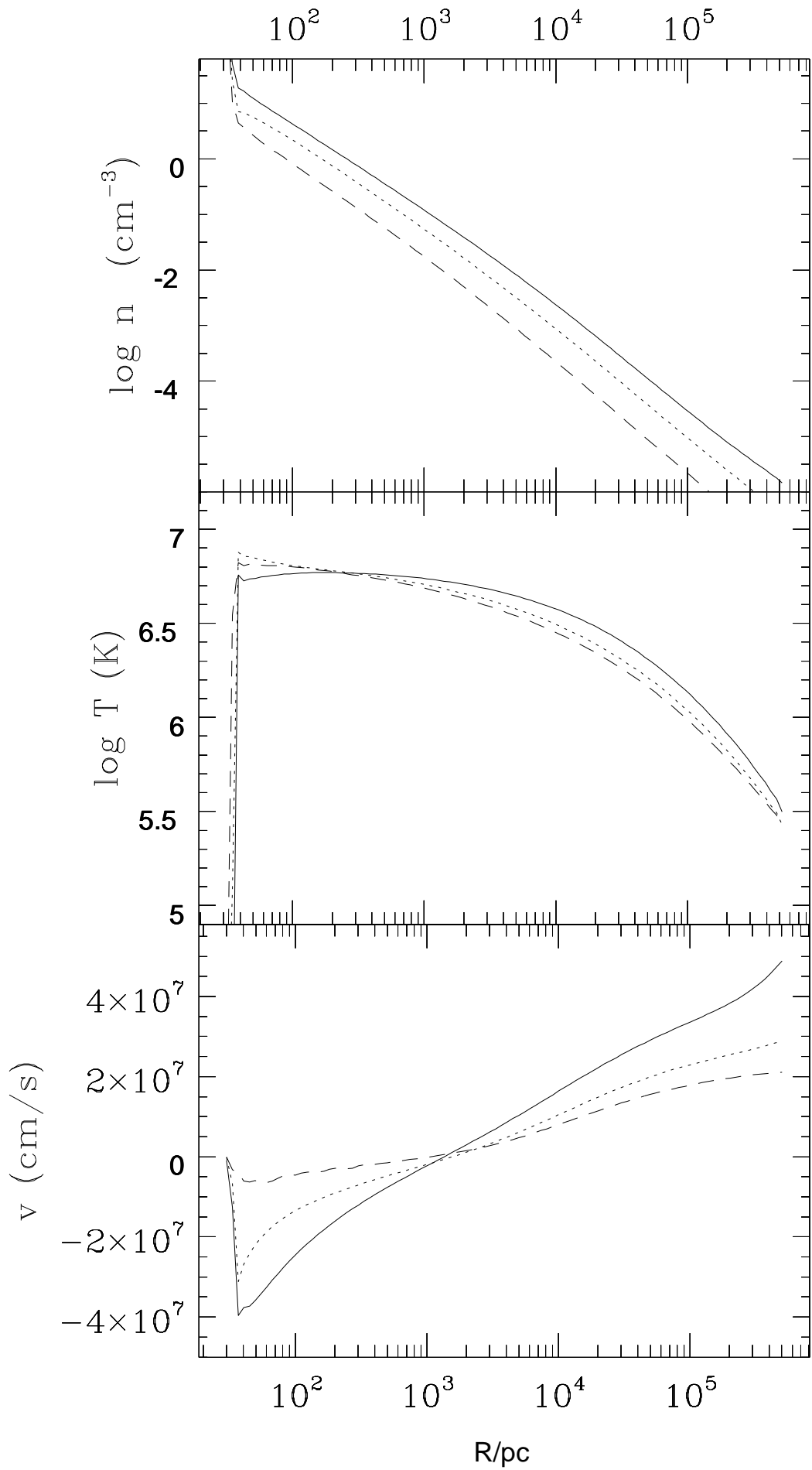
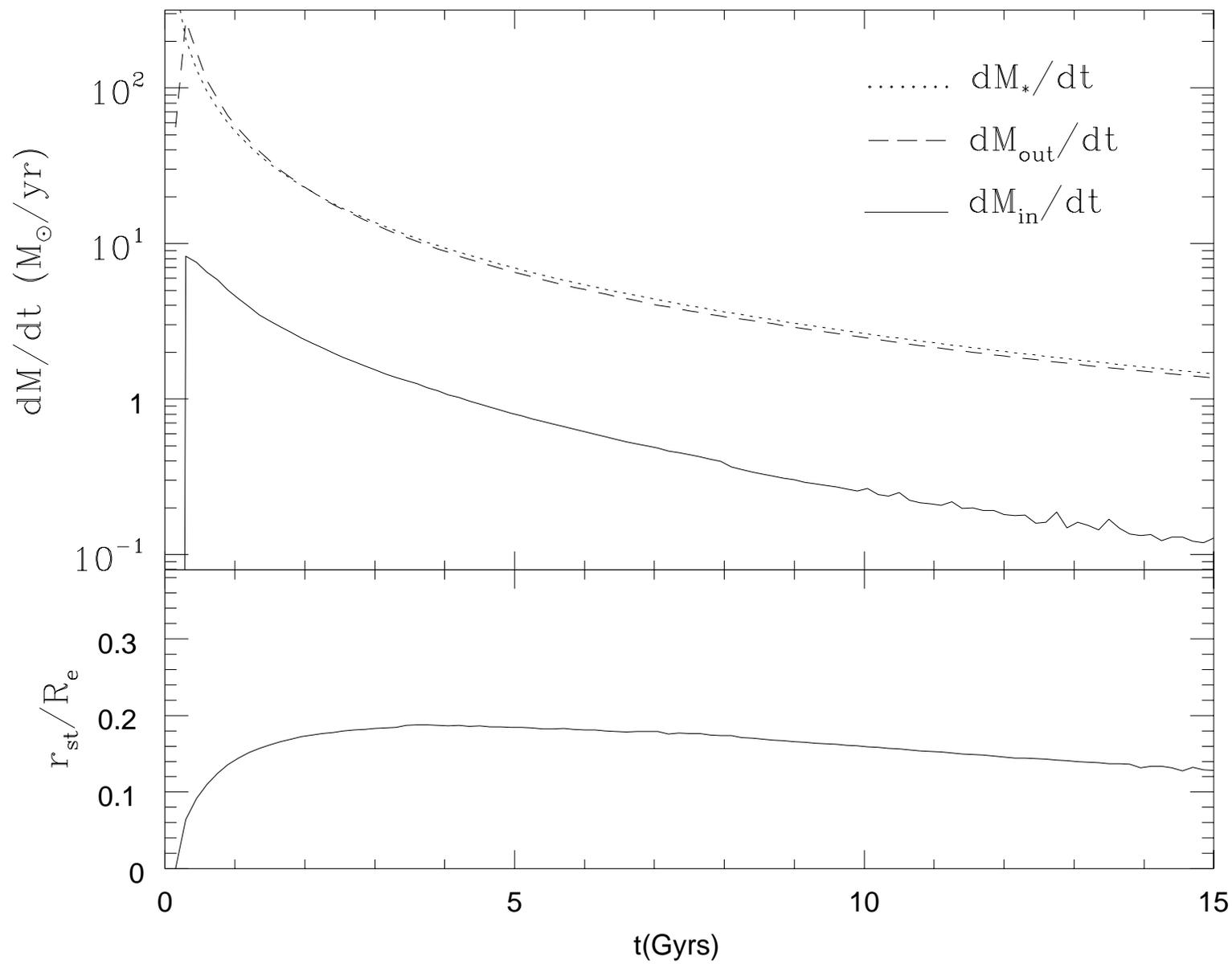
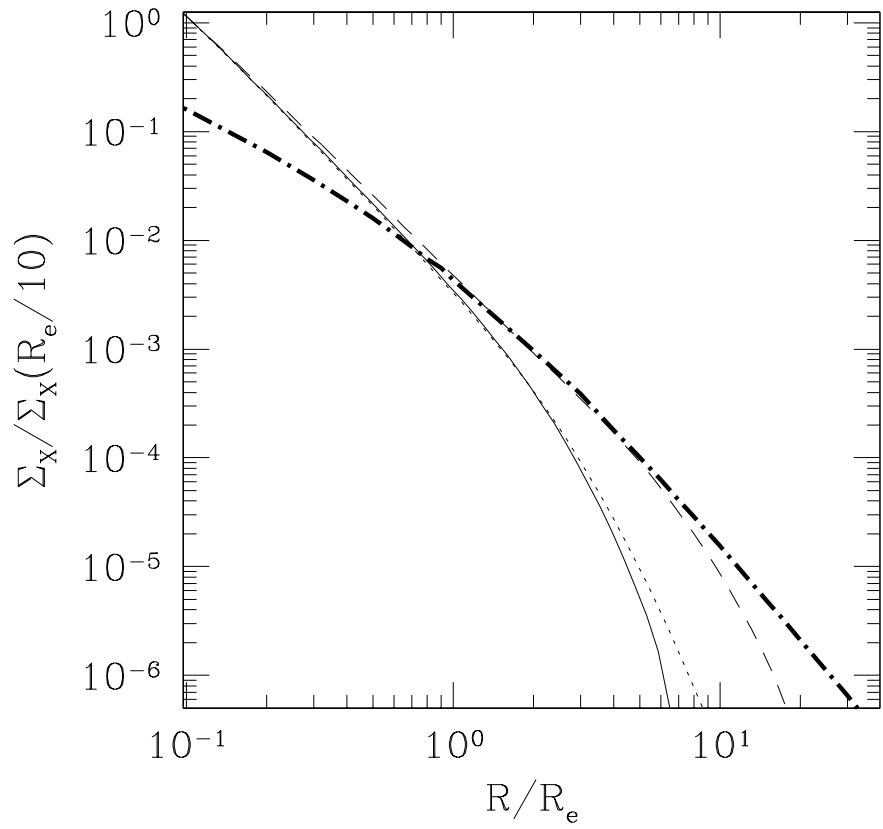
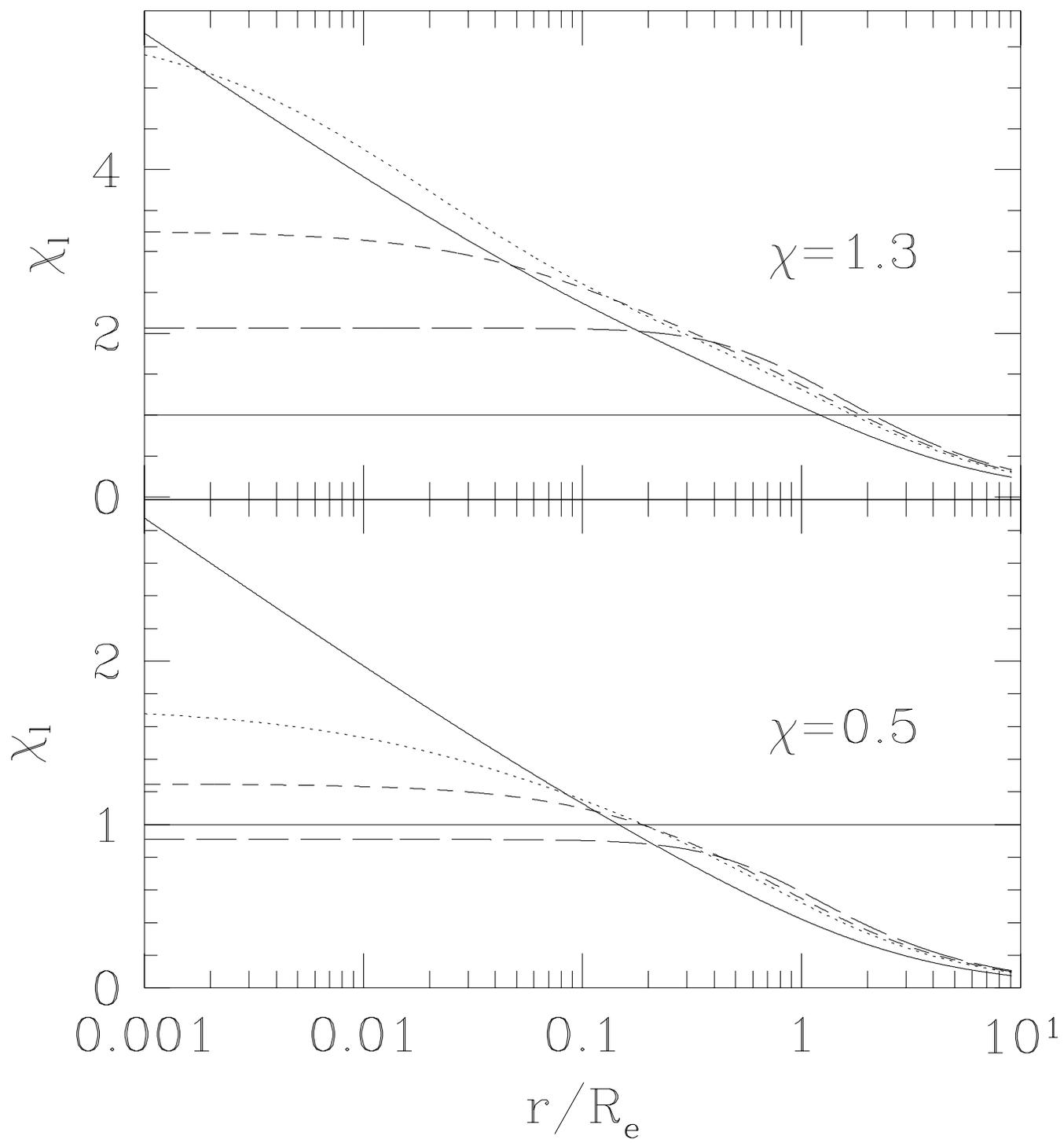
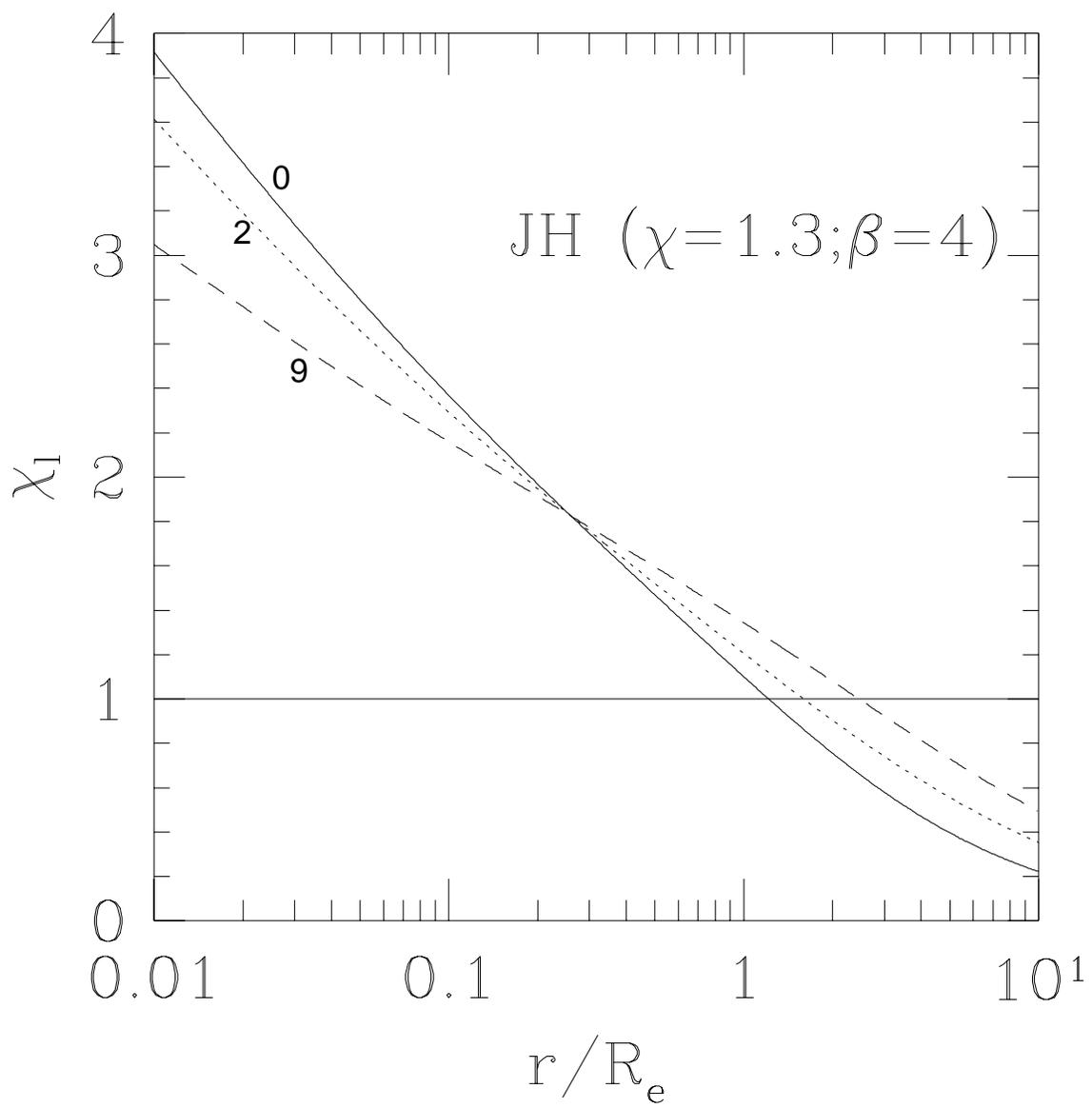
\end{document}